\documentstyle[12pt,epsfig,glas]{article}
\begin{document}
\begin{titlepage}{GLAS--PPE/95--05}{\today}
\title{{
Model of charge transport in Semi-Insulating Undoped GaAs microstrip detectors}}
\author{R.~Bates, S.~D'Auria,
 S.J.~Gowdy, V.~O'Shea, C.~Raine, K.M.~Smith}



\vspace{2cm}

\begin{abstract}
\vspace{0.5cm}
{\small
In this paper we present a method for simulating the response of microstrip 
detectors to minimum ionizing particles, making use of a program for field 
calculation, a program for carrier drift 
and SPICE for circuit response.
A knowledge of the electric field 
 is essential for any further improvement of the program. A simple
model involving EL2 levels in the metastable state
is proposed to explain some measurements of electric field.
}
\end{abstract}

\vspace{2cm}
\centerline{\em Presented by S.~D'Auria at the $4^{th}$ 
workshop on GaAs detectors and
related compounds}
\centerline{\em San Miniato (Italy) 19-21 March 1995}
\newpage
\end{titlepage}
\section{Introduction}
Microstrip detectors made of GaAs will be used in a part of 
the forward tracker detector in
 ATLAS at the 
LHC because of  their larger radiation hardness when compared to silicon 
devices. Test beam results are 
reported elsewere in these proceedings.
In order to optimise the parameters of the devices, such as  total thickness,
strip pitch and aspect ratio for a required 
spatial resolution, we are developing a program which is able 
to calculate the current signal from the various 
geometrical configurations and crystal
properties. A more detailed study will be published later; in the following
 we  describe the method  and stress the importance 
of  a detailed 
knowledge of the electric field inside the detector.

The second part of this paper deals with a model for the electric field.
Semi-Insulating-Undoped (SIU) GaAs has a resistivity of the order of
$10^7 \Omega cm$ which is achieved through a compensation by deep levels.
These levels are mainly of the intrinsic type called EL2, which is due to the 
non perfect stoichiometric ratio between Ga and As. In detectors made of 
this material
the region which is active for detection has
a thickness which increases with bias. At a bias of 200 V about 180$\mu m$
are active \cite{Saverio,OBIC,carlo}.
Various models \cite{Aachen,Freiburg,MacGregor2}
 have been proposed to calculate the electric field, and two 
measurements have been made \cite{craig},\cite{praha}. The measurements 
 show that the electric field is approximately uniform at high voltages,
which means that the sensitive region has a very low  fixed charge density.  
This observation is very difficult to explain in homogeneous materials, where
the fixed charge density is constant over the depth.
A possible explanation of such a field shape is that some of the EL2 defects,
under the influence of a strong electric field,
make a transition to a normally metastable state featuring no energy levels 
in the forbidden gap. The existence of such a level has been reported 
\cite{bourgoin}, but neither evidence nor exclusion of a
voltage-induced transformation has yet been reported.

\section{Microstrip detector simulation}
SIU GaAs is characterized by an imperfect drift of charge carriers, due
probably to trapping \cite{lancaster} and to the peculiar shape of the electric
field. Double-sided microstrip detectors show a marked asymmetry in 
charge response from the two sides \cite{doublesided}.
Therefore a full 2-dimensional simulation of the charge drift in the 
device is needed in order to
calculate the signal induced on the strips.

\begin{figure}
\centerline{\epsfig{file=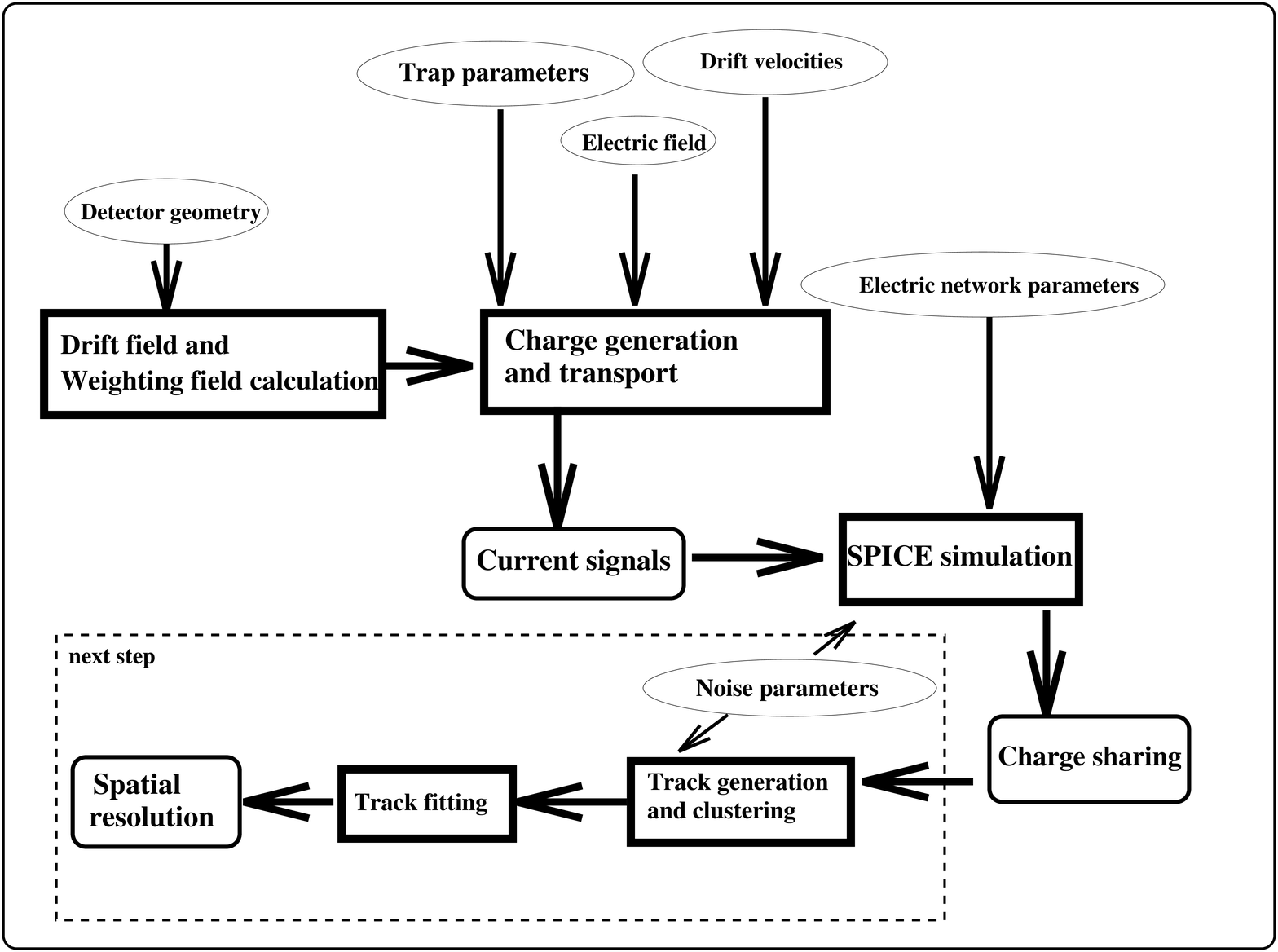,width=3in}}
\caption{\em Block diagram of the codes used for microstrip detector simulation.
Input parameters are in ellipses. \label{fig1}}
\end{figure}

The chain of programs used is shown schematically in fig.~1. The three main 
parts are: a program to calculate the electric fields, one to simulate the 
drift of electrons and holes, and finally a SPICE calculation of the response
of the front-end network. Each of these parts will be considered separately.

\subsection{Field calculation}
In simulating a device two or more electric fields have to be considered 
\cite{gatti}.
 They are 
the physical field which is driving  the motion of charges and the so called 
{\em weighting field} which has dimensions $[L^{-1}]$ and is a purely 
geometrical function of space position. It is a measure of the coupling between
the moving charge and the electrode considered. For instance, in the simple 
case of parallel-plate diode electrodes separated by a distance $d$
the weighting field is a vector of length $1/d$ perpendicular to the 
surface. In the case of devices with more electrodes, 
as in microstrip detectors, the weighting field for the {\em j$^{th}$}
electrode is calculated by holding it at +1~V and mantaining 
 all the other electrodes 
at ground. All the fixed charge has to be neglected.

In the first stage of work reported here
 we have used the code GARFIELD \cite{garfield} 
to calculate the electric fields. This program was written to simulate 
gas operated multiwire drift chambers. A part of it performs a 2-dimensional
calculation of electric field in a configuration where electrodes have a 
wire shape or are infinite planes. 

Each microstrip was simulated as an array
of neighbouring wires with a distance between centres 1\% larger than 
their diameter. We then checked that the resulting field was independent
of the wire diameter and we chose a diameter of $1\mu m$ as a good compromise.
A clear limitation of this method is the lack of simulation of surface
effects with dielectric materials such as silicon nitride and air. Also, we expect
 that results are reliable only on a scale larger than a few times the 
wire diameter. This is acceptable if signals from minimum ionising particles
({\em m.i.p.s}) have to be simulated,
but a more accurate calculation of the field is needed if details have to be 
looked at or if signals from alpha particles are to be simulated.

\begin{figure}
\centerline{\epsfig{file=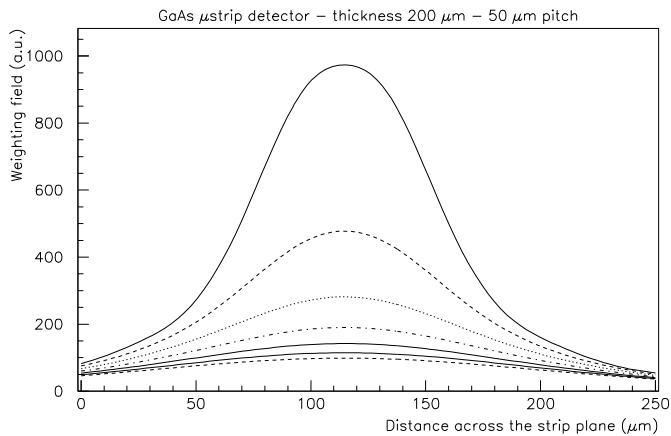,width=3.8in}}
\caption{\em
Modulus of the weighting field for one strip,
 for a $200 \mu m$ thick GaAs microstrip
detector with $50 \mu m$ pitch and $30 \mu m$ metal. The different curves
refer to  distances from
the strip plane differing by 25~$\mu m$.
\label{fig2}}
\end{figure}

A plot of the modulus of the weighting field for one strip is shown in fig.~2.
The major contribution to the signal for that particular strip is due to holes
or electrons moving in proximity of it.  Due to the geometry, however
 most of the signal is due to holes.

\begin{figure}
\centerline{\epsfig{file=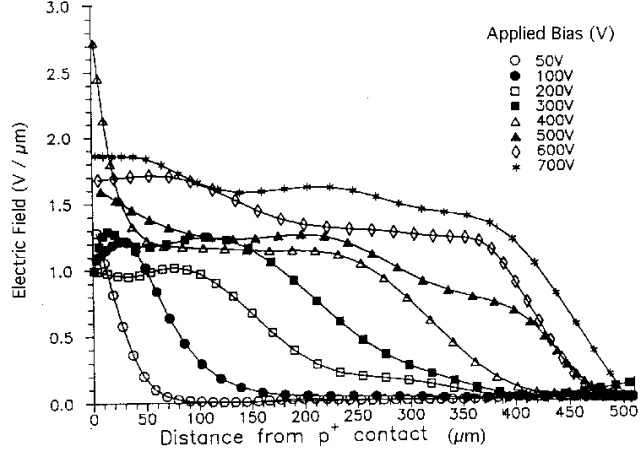,width=3.4in}}
\caption{\em
Modulus of electric field profile  measured by Berwick et al. 
 with a contact
 probe on the cross section of a 500 $\mu m$ -thick SIU-GaAs detector.
\label{fig3}}
\end{figure}

The drift field was calculated by GARFIELD and then multiplied by the following 
function:\begin{equation} f(x) = \frac{1}{1+e^{\frac{x-W}{W_t}}}\end{equation}
where x is the distance from the strip plane, 
$W = W(V)$ is the thickness of the active region, $W_t$ is the thickness
 of the  transition region. A low overall constant field
of 100~V/cm was then added. The result  is 
an approximation to the field  measured in ref.~\cite{craig}, (fig.~3). 
It was assumed that $W = \alpha V$ with
$\alpha = 1 \mu m/V$, as reported in \cite{OBIC}.

As an example, a zone of 550~$\mu m$ in 
width for a 200~$\mu m$ thick detector was simulated. The 
strip pitch was 50 $\mu m$ and the metal width was 40~$\mu m$.
Only the part comprising  5 central strips was used for further analysis
corresponding to a grid of $ 90 \times 90$ points.
Interpolation of the electric field was used between points.

\subsection{Drift of charge carriers}
The initial distribution of charge is assumed to be uniform along the
linear track of a minimum ionizing particle. Genaration of delta rays and
charge spreading has not yet been implemented, 
but is assumed to be not essential at the
present stage. A constant number of electron-hole pairs is generated for each
track. A time-slice approach is used in simulation, as a current signal is
required. Each individual carrier is followed during its drift. 
Trapping probabilities are calculated from the trap concentrations , $N_k$,
 cross sections $\sigma_k$ and drift velocity $v(x(t))$. A
 flag   associated with each carrier indicates its status, which can
be: free, trapped, reached the electrode.
The probability of being trapped by the {\em k$^{th}$} type of trap 
is given by: \begin{equation} 
P_k (\sigma_k, N_k, x, \Delta t) = 1 - exp{(- v(x) \sigma_k N_k \Delta t})
\end{equation}
where $\Delta t$ is the time step.

\begin{figure}
\centerline{\epsfig{file=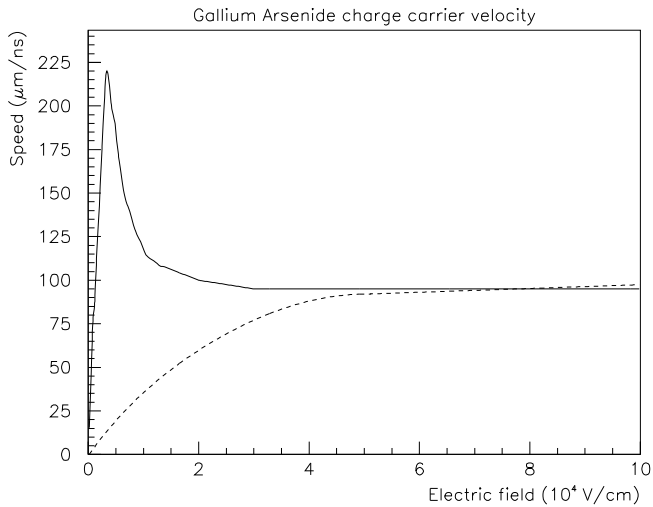,width=4in}}
\caption{\em
Electron and hole drift speed used in the program. The saturation velocity is
nearly the same for electrons and holes in this material.
\label{fig4}}
\end{figure}

 Detrapping is also taken into account as a possible process. 
We assume that there is 
a mean
trapped time $\tau_k$ and that both $\tau_k$ and $\sigma_k$ are independent 
of electric field. 

The carrier drift velocity is calculated from the electric field using 
 parametrizations of the experimental data 
\cite{kino}\cite{dalal}, as shown in fig.~4. 
We note that provided the field is high enough, the details of its shape
have a negligible influence on carrier velocity for the larger part of the 
device, as saturation is reached.  If drift in a magnetic field and 
diffusion have to be considered, however,
an exact knowledge of the electric field is 
essential. 

\begin{figure}
\centerline{\epsfig{file=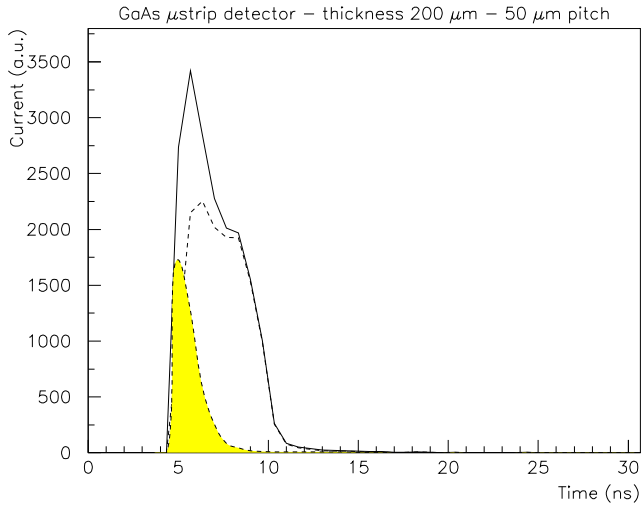,width=4in}}
\caption{\em
An example of the simulated current signal generated by minimum ionizing particles
traversing a $200 \mu m$ thick detector in the middle of the considered strip,
which in this case is $30 \mu m$ wide. 
The bias voltage was 180~V. The major contribution is due to holes
which approach the area where the weighting field is high. The electron 
contribution is shown in the shaded curve.
\label{fig5}}
\end{figure}

The current signal on the {\em j$^{th}$} electrode 
is calculated from the Ramo-Shockley theorem:
\begin{equation} i_j(t) =  \sum_{n=1}^{N_c} q_n \vec{v}_n(t) 
\vec{E}_{w_j}(\vec{x}_n(t)) \end{equation}
where $N_c$ is twice the number of electron-hole pairs generated,
$\vec{E}_{w_j}(\vec{x}_n(t))$ is the weighting field and $q_n$ the  charge of the 
carrier.
The charge signal is obtained by integration of $i(t)$ and gives the charge 
collection efficiency.
An example of a current signal is shown in fig.~5 and is 
used as input for SPICE.

\subsection{SPICE calculation}

\begin{figure}
\centerline{\epsfig{file=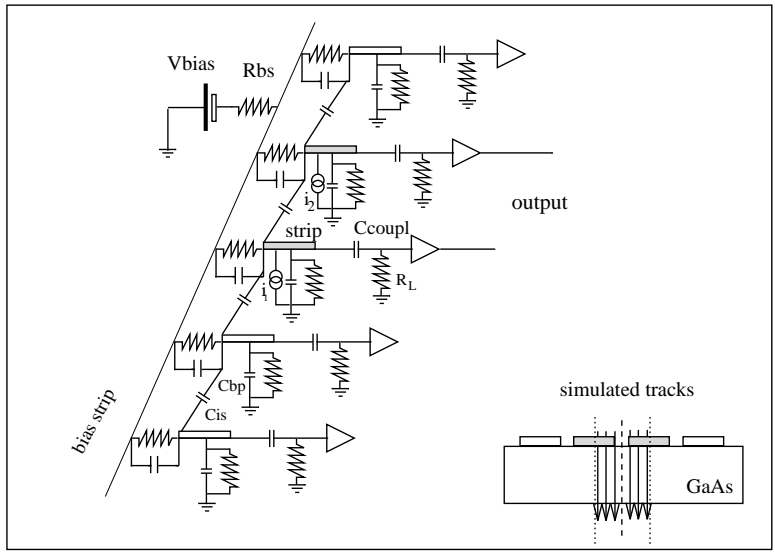,width=4in}}
\caption{\em
Schematic of the circuit used in the SPICE simulation. C$_{is}$ is the 
inter-strip capacitance, C$_{bp}$ is the capacitance to the backplane, 
$R_L$ is the load resistance of the amplifier. The biasing structure is 
represented by a parallel RC. Current sources from two neighbouring strips
are taken into account. An array of seven strips was simulated.
\label{fig6}}
\end{figure}

In order to calculate the signal reaching each preamplifier the front-end and
bias coupling network was simulated using SPICE. The parameters depend mainly
on the detector geometry. The main parameters are the inter-strip capacitance,
depending on pitch, aspect  ratio and strip length, the strip to back contact
 capacitance and the decoupling capacitance. The biasing structure, based on 
punch-through was also included as a 10~G$\Omega$ resistor with a 20~fF 
capacitance in parallel. A schematic is shown in fig.~6. Resistors had to be 
used to ensure  a DC path to nodes. 

The signals from two neighbouring strips were included as two independent 
piece-wise linear current generators. The current values were calculated with 
the drift code described previously, simulating tracks impinging 
perpendicularly on the strips; the hit position was varied from the centre of
the central strip to the centre of the neighbouring one on the right.
It was assumed that all strips were read-out.
In simulating the signal a symmetry relation was assumed, i.e. 
$ i_j(t)|_{y=y_0} = i_{j+1}(t)|_{y=p-y_0} $ where $p$ is the pitch, and $y_0$
is the distance of the track from the centre of the $j^{th}$ strip.

\begin{figure}
\centerline{\epsfig{file=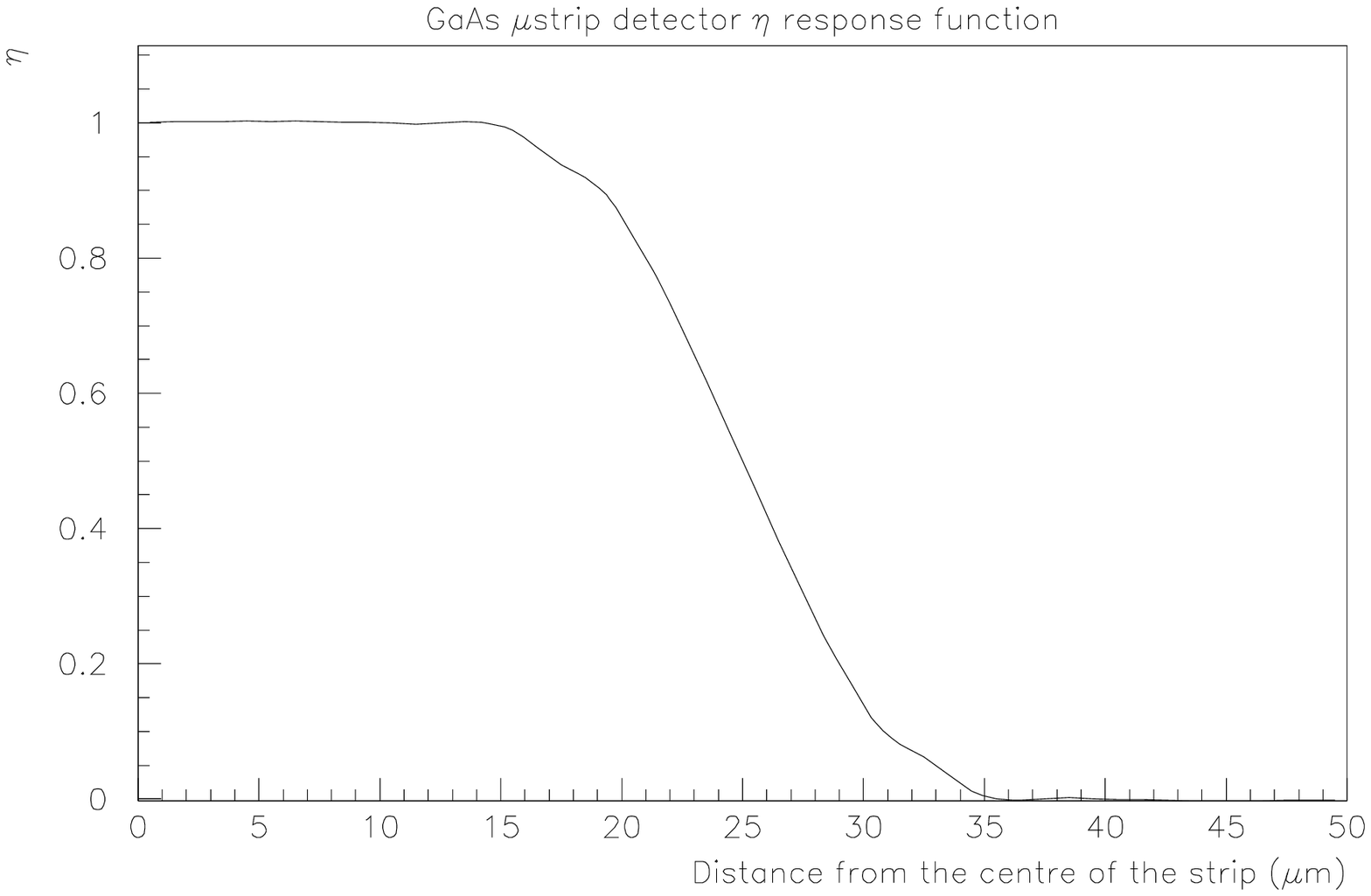,width=3.8in}}
\caption{\em
Plot of the $\eta$ variable defined as: 
$ \eta(x)=\frac{Q_{left}(x)}{Q_{left}(x)+Q_{right}(x)}$,
where $x$ is the distance from the centre of the strip. The detector is 
 $200 \mu m$ thick, pitch is
 $50 \mu m$, metal width is  $30 \mu m$. \label{fig7}}
\end{figure}

The current signal at the preamplifier input  was calculated and then 
integrated to give the total charge detected by each channel. An $\eta$ 
function \cite{turchetta},
 related directly to spatial resolution
could be calculated for the given configuration. An example is shown in fig.~7.

\section{A model for the electric field}
The present simulation chain can be a starting kernel for further improvements.
However, a more detailed simulation, including effects
 of magnetic field and diffusion,
is possible if a definite shape of electric field is generally accepted. 
Measurements made at UMIST  \cite{craig} indicate that
the field is uniform
in the vicinity of the reverse biased contact, as shown in fig~3. 
This a clear  experimental result. Most calculations \cite{Aachen}
\cite{Freiburg} result instead in a linearly decreasing electric field.
 This arises naturally if a uniform doping concentration is assumed. Although 
some of the defects are not ionized due to quasi-Fermi-level shifting by the
reverse bias 
current, the main trend is a field decreasing monotonically to the
back contact. A constant electric field is quite difficult to explain.
 A correlation between reverse current and 
charge collection efficiency for alphas has been reported 
as evidence  supporting  the calculations. This correlation has not been 
found by us for {\em m.i.p.s}, as
we have several examples of detectors with a low reverse current density 
(20 nA/mm$^2$) and high charge collection efficiency (90~\%) 
over a large thickness
(500~$\mu m$) (fig.~8).

Another model \cite{MacGregor2} proposes a neutralization of deep levels
by field-enhanced trapping of the electrons of the reverse current.
In this model, too,  a high current is required to obtain a
 high charge collection efficiency.

The nature of the reverse current is not yet clear. Thermionic emission would 
require a really low barrier height, while if a generation mechanism is 
assumed,
 saturation could normally be explained only 
if  most  of the current is generated in
the first few micrometers below the reverse-biased contact. 
Otherwise the current density is expected to increase with the volume in which
 the  electric field  is high.

\begin{figure}[t]
\centerline{\epsfig{file=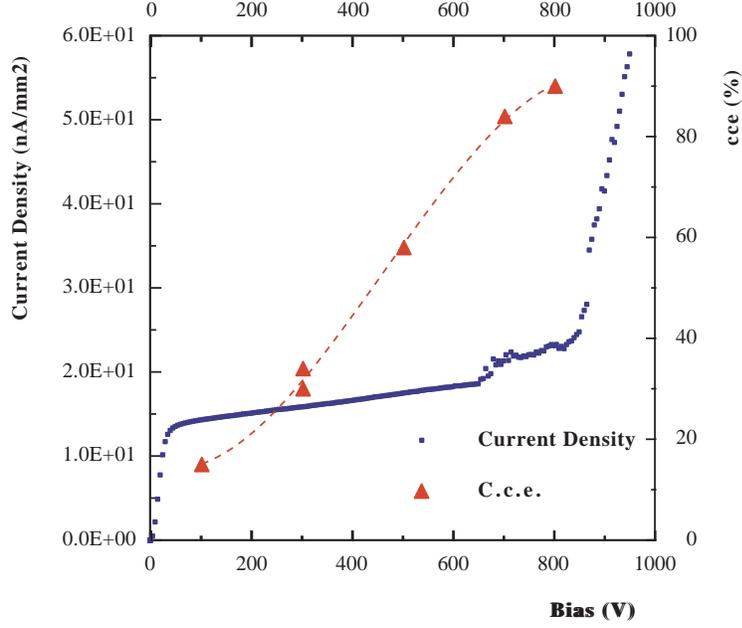,width=4in}}
\caption{\em
Charge collection efficiency 
for {\em m.i.p.s} and leakage current density at 20.0$^o$~C 
for the SIU-GaAs pad detector G114-5.
The thickness is 500~$\mu m$, area is 7~mm$^2$, with guard-ring.
\label{fig8}}
\end{figure}

Commercial SIU-GaAs usually contains a concentration of about 
$10^{16}$~cm$^{-3}$
of EL2 defects. This defect
 is believed to be due to an As substitutional 
($As_{Ga}$) plus
an As interstitial ($As_{i}$) as second nearest neighbour \cite{bourgoign-crit}.
The material is naturally compensated and shows a resistivity of $10^7 \Omega$cm
which is ideal in a substrate for electronic and optical circuits. 

The EL2 defect has been reported \cite{bourgoin} to have a metastable state
EL2$^*$ which has  very little or no electrical and optical
activity.  It is therefore believed that the corresponding energy level is
split and shifted into the conduction and valence bands.
One proposed microscopic structure of EL2$^*$ is similar to EL2, but with the
$As_i$ atom as nearest neighbour. 
the transition is: \begin{equation} EL2 \rightarrow EL2^*   \Longrightarrow
As_{i(II)} \rightarrow As_{i(I)}\end{equation}

The EL2$^*$ can be generated via optical
absorption, involving different charge states \cite{kiliulis}.
The decay rate to the stable state has an activation energy of 0.3~eV so that 
at room temperature the metastable state has a lifetime of about $1.5~\mu s$.
It is generally observed at fairly low temperatures, (about $77 K$) and 
in the presence of 
a low electric field. 

If the transition could happen at room temperature, under the effect of a high
electric field, for example,
 the measurements of electric field shape in \cite{craig} 
could be qualitatively explained in the following way:
as reverse bias is applied, deep donors start to become ionized;
 some  are
neutralized by trapping electrons from the leakage current, as in 
\cite{MacGregor2}; this lowers their effective concentration, in such a way 
that the electric field can extend for several micrometres. This process 
continues until the field at the contact reaches a critical field, $E_c$,
of about 10~kV/cm, at a bias voltage between 20 and 50~V.
At this point the transition 
EL2 $\rightarrow $EL2$^*$ starts to take place, so that the field 
remains roughly constant, as deep levels are no loger active. 
We should assume 
that the transition is not complete, leaving some compensating EL2$^+$.
The region where the field drops is shifted rigidly towards the back 
contact as bias
is increased. In this transition 
region the electric field is non-zero and most of the
EL2 defects are active. Thus the 
reverse current can be efficiently generated there and
is constant to the extent that the transition region has a 
 constant thickness while
moving across. This is to be expected in a homogeneous material.
A sketch of the proposed model is shown in fig.~9.

\begin{figure}
\centerline{\epsfig{file=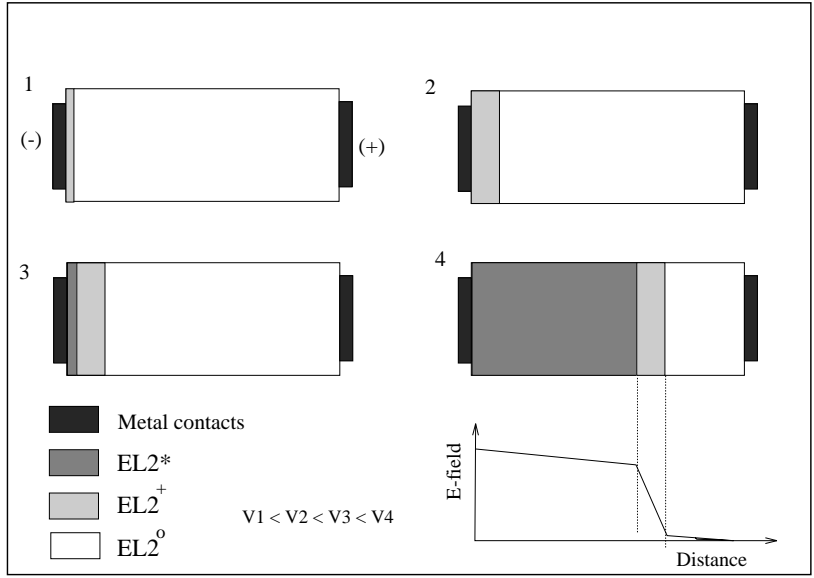,height=3.in}}
\caption{\em
Cross sections of a SIU-GaAs detector at different reverse bias voltages,
showing the regions
 with a different predominant EL2 state.
\label{fig9} }
\end{figure}  

From the point of view of charge transport, three regions of electric field
can therefore be distinguished:
one  constant, one linearly decreasing and one with an
extremely low residual electric field. 
This is quite similar to that shown in fig.~3.
In the third region transport is mainly due to charge relaxation, 
with a typical time of
$\epsilon \rho \approx 10 \mu s$, giving a slow signal component.
In the transition region the field is decreasing. Thus the electron 
velocity  initially  increases  then decreases and trapping can be enhanced.
The charge collection efficiency and  
the medium-slow components of the signal depend on the shape of the field. 
In the region where a constant field is present the charge transport is 
good, as the concentration of active traps is lower. Little contribution to 
trapping is expected from this zone.

The shape of electric field used in the simulation now has  also a theoretical
(although qualitative) explanation and is not based only on empirical data.
While the validity of this model is not proved, it enables to explain 
consistently many features of  SIU-GaAs detectors. 
{\em C-V} characteristics would not characterise this model, as the transition region
where charge can move has the same features as in normal semiconductors.
The saturation of reverse current can be explained in this model, assuming that
the current is generated mainly in the transition region where the electric
field is non negligible and the concentration of active levels is high.  

A possible way to test the model is with OBIC measurements 
\cite{OBIC} using a focused light beam 
at a wavelength which is unable to separate directly an
electron-hole pair. In this way only deep levels will be probed and the
photocurrent at a distance from the front contact
will be proportional to the number of active levels at that point and to the 
electric field. A region of constant width where  the photocurrent  is higher
is expected to move across
 the detector as the reverse bias is increased. If the model is incorrect
 the region with high
photoresponse will simply increase in width.
We propose to carry out this measurement in the near future.

\section{Conclusions}
A  code has been set up to simulate the response of GaAs microstrip detectors
to {\em m.i.p.s}.  From preliminary results it is evident that charge 
sharing is  possible in principle also in SIU-GaAs detectors where the 
 charge transport is not perfect, provided that the 
strip parameters and noise performances
are optimized. 
 More extensive results on various geometry and parameters will be
presented soon.
Limitations of the present code are in the calculation of details of
the electric field near the strip and at the  
interface with passivating layers. This will be improved 
 using a more suitable package for the field calculation.
Further developments will also
 aim at including the possibility of drift in a
magnetic field and the effects of diffusion. Because of the peculiar band 
structure of GaAs these processes are strongly dependent on the local intensity
of the electric field even when
the drift velocity is saturated.
A detailed knowledge of the electric field is needed to improve the simulation.

A possible explanation of  the  measured 
electric field profile has been proposed,
based on a transition of the EL2 defects which makes them electrically
and optically inactive. 
This model can explain consistently a variety of features
of SIU-GaAs, like the slow transients and a saturation in generation
current.

\section*{Acknowledgements}
We wish to thank PPARC and the EC 
(via an H.C.M. contract) for support and our host for organizing this
workshop.

\end{document}